\documentclass[a4paper,11pt]{article}
\usepackage{pos}

\let\OLDthebibliography\thebibliography
\renewcommand\thebibliography[1]{
  \OLDthebibliography{#1}
  \setlength{\parskip}{0pt}
  \setlength{\itemsep}{0pt plus 0.3ex}
}

\title{Reconstruction of extensive air shower images of the first Large Size Telescope prototype of CTA using a novel likelihood technique}
 \ShortTitle{Reconstruction of EAS images of the LST-1 of CTA using a novel likelihood technique}

\author*[a]{Gabriel Emery}
\author[a]{Cyril Alispach}
\author[a]{Mykhailo Dalchenko}
\author[a]{Luca Foffano}
\author[a]{Matthieu Heller}
\author[a]{Teresa Montaruli}

\affiliation[a]{ University of Geneva - DPNC,\\
  24 rue du G\'{e}n\'{e}ral-Dufour, Geneva, Switzerland}

\forColl{CTA LST project} % W/O "Collaboration"

\emailAdd{Gabriel.Emery@unige.ch}
\emailAdd{cyril.alispach@gmail.com}
\emailAdd{Mykhailo.Dalchenko@unige.ch}
\emailAdd{Luca.Foffano@unige.ch}
\emailAdd{Matthieu.Heller@unige.ch}
\emailAdd{Teresa.Montaruli@unige.ch}

\abstract{Ground-based gamma-ray astronomy aims at reconstructing the energy and direction of gamma rays from the extensive air showers they initiate in the atmosphere. Imaging Atmospheric Cherenkov Telescopes (IACT) collect the Cherenkov light induced by secondary charged particles in extensive air showers (EAS), creating an image of the shower in a camera positioned in the focal plane of optical systems. This image is used to evaluate the type, energy and arrival direction of the primary particle that initiated the shower. This contribution shows the results of a novel reconstruction method based on likelihood maximization. The novelty with respect to previous likelihood reconstruction methods lies in the definition of a likelihood per single camera pixel, accounting not only for the total measured charge, but also for its development over time. This leads to more precise reconstruction of shower images.
The method is applied to observations of the Crab Nebula acquired with  the Large Size Telescope prototype (LST-1) deployed at the northern site of the Cherenkov Telescope Array.}

\FullConference{37$^{\rm{th}}$ International Cosmic Ray Conference (ICRC 2021)\\
		July 12th -- 23rd, 2021\\
		Online -- Berlin, Germany}

\begin{document}
\maketitle

\section{Introduction}
Ground-based observations in the field of very-high-energy (VHE) gamma-ray astrophysics deal with the detection of energetic photons in the energy range between O(10) GeV and O(100) TeV from cosmic sources. VHE gamma-ray fluxes interact in the atmosphere, producing electromagnetic showers detectable over large areas from the ground. Secondary charged particles in these extensive air showers (EASs) produce Cherenkov light which can be collected by Imaging Atmospheric Cherenkov Telescopes (IACTs).
The challenge of IACT-based observations is two-fold. Firstly, gamma rays are overwhelmed by an about $10^5$ orders of magnitude larger background due to cosmic rays.
Their hadronic nature can be inferred due to their more sparse images compared to the elongated images of gamma rays. In addition, triggered images are contaminated by the Night Sky Background (NSB), which includes light from the stars and terrestrial sources. Secondly, the primary particle information, namely its energy and direction of origin, needs to be inferred from the sampled image of the EAS.

The Cherenkov Telescope Array (CTA) will be a new observatory aiming at observing the VHE sky using two arrays of IACTs. Located in the northern and southern hemispheres for a near full sky coverage, the arrays will be composed of telescopes with a total of three different sizes. The different sizes of mirrors make it possible to cover an extended energy range compared to current instruments. With largely improved performance exploiting the experience and technological development from the current generation of IACTs, CTA will be a valuable source of information about the most energetic phenomena in the Universe in the decades to come.
The Large Size Telescope prototype (LST-1) is the first of four LSTs to be constructed in the northern site of CTA. The LSTs will be most sensitive in the lower side of CTA energy range down to about 20~GeV, hence providing an overlap with the space-borne gamma-ray telescopes. LST-1 was inaugurated in October 2018 and it is now in commissioning phase and already taking data.

In this work, a novel likelihood maximization technique, exploiting the space-time development of EAS images recorded by IACTs of CTA is introduced. It was first developed for an IACT camera using silicon photo-multipliers (SiPM)~\cite{Cyrilthesis}. Here we adopt and apply it to  the LST-1 prototype, whose camera uses photo-multiplier tubes (PMTs). The method is applied both to simulated and to observed data from the Crab Nebula.

\section{Image likelihood}

The reconstruction of the properties of the shower-initiating particle using images from Cherenkov telescopes is a complex problem for which a variety of methods have been developed. Most modern techniques use the distributions of integrated charges in each pixel, sometimes adding information on the time of maximum signal in each of them. The most well-known and applied method, introduced by Hillas~\cite{hillas_parameters}, derives a set of parameters (generally named as Hillas parameters and sometime slightly extended compared to the original set) describing each EAS image. These parameters are then used to recover primary particle properties. A reconstruction method developed for the LST chain of data analysis uses random forest regressors and a classifier for this purpose. Additionally, a variety of machine learning based methods are used in IACT analysis (see, e.g.~\cite{2021APh...12902579S,vuillaume:hal-02197396}). Another likelihood method is presented in Ref.~\cite{2009APh....32..231D}. It finds the shower parameters by minimising a likelihood function over the integrated charge in the pixels from a database of gamma events image templates. The method uses the distribution of the integrated charge in each pixel, while our method uses the recorded waveforms and the detailed response of the photosensor. A likelihood is evaluated in each pixel for each recorded time, and the total image likelihood is then the product of all these terms. The signal recorded for one pixel and one time interval will be called a sample in the following. The different single sample likelihoods are linked by a space-time model including predicted image properties and instrument response. The fitting procedure is performed by maximizing the total likelihood over the parameters of the image model. The large parameter set leads to a non-negligible computation time which should be accounted for in future optimisations.
\newline

\textbf{Single sample likelihood : }
The single sample likelihood can be divided into two parts. The first part accounts for the statistical fluctuations in the number of photo-electrons $k$ detected in a pixel compared to the average expectation value $\mu$ from an EAS. As the method was originally developed for a SiPM-based camera, a generalised Poisson distribution accounting for cross-talk was initially used~\cite{2012NIMPA.695..247V}. For the case of LST PMTs, we assume negligible cross-talk ($\mu_X = 0$). The second part of the likelihood evaluates the probability of measuring a charge $W$ knowing $k$, the temporal development of the response of the pixel receiving a photo-electron $T$ and the time at which it is evaluated $\Delta t$, the gain $G$, the baseline $B$ corresponding to the average pixel response when no signal is received, the baseline fluctuation width $\sigma_e$ and the single photo-electron charge response width $\sigma_s$. The single sample likelihood is thus obtained by summing over each possible value of $k$ as follows :
\begin{equation}
    L(W | \mu, G, B, T, \Delta t, \sigma_e, \sigma_s) = \sum_{k=0}^{+\infty} \frac{\mu^{k}}{k!}e^{-\mu} \times \frac{1}{\sqrt{2\pi}\sigma_k}exp(-\frac{(W - kGT(\Delta t) - B)^2}{2\sigma_k^2})
\end{equation}
where $\sigma_k^2 = \sigma_e^2 + k \sigma_s^2$.
\newline

\begin{figure}
    \centering
    \includegraphics[width = \textwidth]{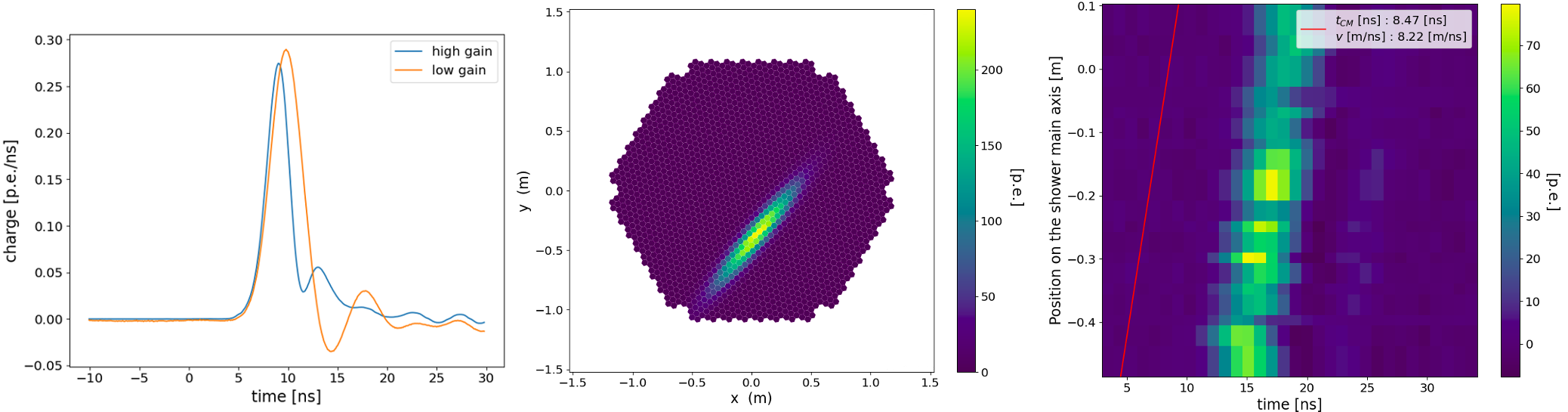}
    \caption{\textit{left}~: Template of the pulsed response of a pixel to a single p.e. \textit{middle}~: 2D Gaussian spatial model as fitted to a MC gamma event. \textit{right}~: Signal distribution as a function of time and of the position along the fitted main axis on the same event. The red line is the fitted linear shift between the time of arrival of the signal at different position along the shower main axis.}
    \label{fig:model}
\end{figure}

\textbf{Space-time image model : }
To construct the EAS image model we have to take into account the dynamics of the shower development and the model of the instrument response. The latter predicts the evolution of the signal observed in a pixel which is represented by a pulse template. In LST-1 two gain channels are available (named high and low gain). Thus two templates are introduced (Fig.\ref{fig:model}-\textit{left}).

The shower development is represented as a convolution of a 2D Gaussian spatial model for the integrated charge distribution (Fig.\ref{fig:model}-\textit{middle}) and a linear temporal model for the arrival time of the signal in each pixel along the spatial model main axis (Fig.\ref{fig:model}-\textit{right}). Additionally, no dispersion of the arrival time in a single pixel is included. Both the implementation of a more complex spatial model and of the photon arrival time dispersion are under study. 
Currently, the spatial model is characterised by a set of 6 parameters : the normalisation $I$, the position of the shower's center of gravity in the camera frame $(x_{CM}, y_{CM})$, the Gaussian standard deviation along its main and secondary axis $(\sigma_l, \sigma_w)$ called the length and width, and the angle $\psi$ between the shower main axis and the camera $x$ axis. The spatial model predicts the expectation value of the integrated charge in each pixel $\mu_i$. The subscript \textit{i} will be used to identify the pixel at position $(x_i, y_i)$ while the subscript \textit{j} will be used for times.
\begin{equation}
    \mu_i = I \frac{1}{\sqrt{2\pi}\sigma_l}e^{-\frac{1}{2}\frac{l_i^2}{\sigma_l^2}} \frac{1}{\sqrt{2\pi}\sigma_w}e^{-\frac{1}{2}\frac{w_i^2}{\sigma_w^2}}
\end{equation}
    where : 
\begin{equation}
    l_i = (x_i - x_{CM}) cos\psi + (y_i - y_{CM}) sin\psi \; \; \text{and} \; \;
    w_i = -(x_i - x_{CM}) sin\psi + (y_i - y_{CM}) cos\psi
\end{equation}
The temporal model is characterised by two parameters : the time gradient $ \rm v^{-1}$ representing the linear coefficient between the time of the signal development in each pixel along the projected position on the shower main axis, and $t_{CM}$ the reference time of the time development at the spatial model center. The temporal model predicts for each pixel a reference time $t_i$ corresponding to the time 0 of the pulse template:
\begin{equation}
    t_i = t_{CM} +  l_i\ \rm v^{-1} .
\end{equation}

\textbf{Likelihood fitting procedure : }
The total image log-likelihood is given by summing over pixels and recording times. Additionally we introduce a weight $w_{ij} = 1 + |(W_{ij}-B_i)/\text{max}(W-B)|$, in order to increase the importance of samples with high signal compared to those close to the baseline. Before normalisation by the total weight the total image log-likelihood is given by:

\begin{equation}
  \begin{aligned}
    lnL_{tot} & = \sum_i \sum_j w_{ij} lnL\\
      & = \sum_i \sum_j w_{ij} ln \sum_{k=0}^{+\infty} \frac{\mu_i ^{k}}{k!}e^{-\mu_i} \times \frac{1}{\sqrt{2\pi}\sigma_{ki}}exp(-\frac{(W_{ij} - kG_iT_i(t_j - t_i) - B_i)^2}{2\sigma_{ki}^2})
  \end{aligned}
\end{equation}

Two approximations are added in order to obtain a finite and reasonable computation duration. In pixels with low luminosity, namely pixels for which the Poisson term in $L$ is below $10^{-6}$ for $k > 200$,  the sum is limited to values of $k$ for which the Poisson term is above $10^{-6}$.
In the other pixels, a Gaussian approximation to the single sample likelihood is applied. In this case the likelihood is given by :
\begin{equation}
    L(W | \mu, G, B, T, \Delta t, \sigma_e, \sigma_s) = \frac{1}{\sqrt{2\pi(\sigma_e^2 + \mu (GT(\Delta t))^2)}}exp(-\frac{(W - \mu GT(\Delta t) - B)^2}{2(\sigma_e^2 + \mu (GT(\Delta t))^2)})
\end{equation}

Prior to the fitting of the likelihood model, we perform the standard Hillas parameters extraction, extending it with additional temporal features. It is used to provide the initial seed for the model parameters and assign the bounds on the parameter space. A reduction of the pixels and temporal window used in the likelihood is also performed at this point. This improves the fit stability and computation duration using the migrad method of iminuit~\cite{iminuit} to minimize $-2lnL_{tot}$.

\section{Reconstruction of primary parameters}

As mentioned above, the goal of the reconstruction methods of IACTs is to determine primary particle type, energy and direction. With the analytical image model used in this work, the energy and direction of arrival of the primary particle are not directly reconstructed. It could be implemented either analytically, if reliable relations between the primary particle properties and the image can be derived, or through a pre-tabulated image database, as done in Ref.~\cite{2009APh....32..231D}.

The solution adopted in this work is to use a simple machine learning approach using two random forest regressors trained on Monte Carlo simulation to extract the energy and direction of origin of the primary particle. Additionally, a random forest classifier is also used to separate gamma-like images from hadron-like images.
The random forest is trained on diffuse gamma and proton simulations assuming observations with a telescope pointing at $20^\circ$ from the zenith. Parameters used are the space-time model parameters, the second moments from the Hillas parameters extraction and the leakage information. The classifier also uses the regressor outputs. 

\section{Performance on LST-1 simulations}

In this section, we present the performance of the method for the source independent analysis on a Crab Nebula like source. MC simulation of diffuse electrons, protons and on source gammas simulated at an offset of $0.4^\circ$ from the camera center are considered. The MC assumes observations with a telescope pointing at $20^\circ$ from the zenith. The electron and proton distributions are weighted to the spectra from Ref~\cite{Zyla:2020zbs}. For the gamma-ray simulation the Crab Nebula spectrum used is from Ref.~\cite{2004ApJ...614..897A}.

\begin{figure*}[h]
  \centering
  \includegraphics[width=\textwidth]{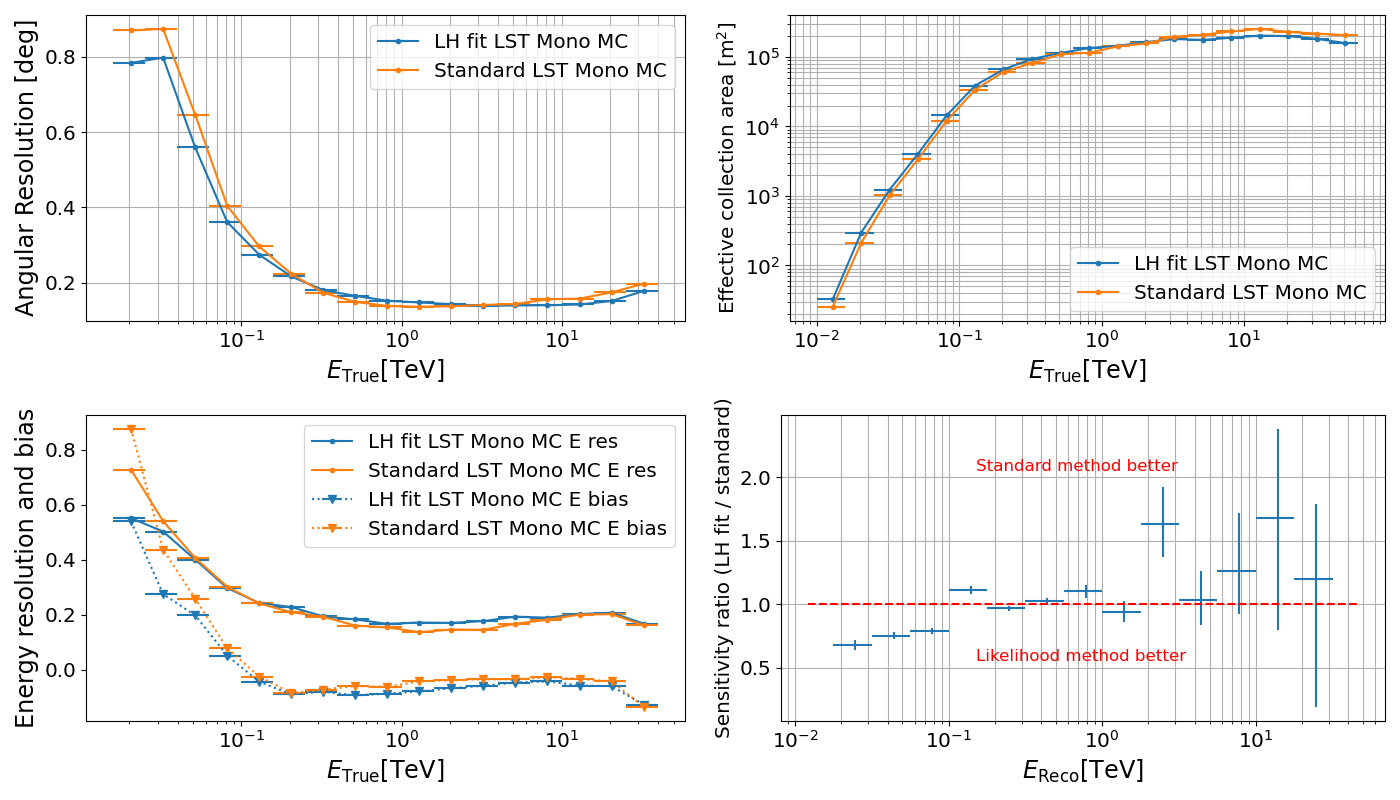}
  \caption{Performance of the method evaluated on MC and comparison with performance of the same analysis procedure with the standard reconstruction. \textit{top left}: Angular resolution : 68\% containment angle. \textit{top right}: Effective collection area. \textit{bottom left}: Energy resolution : 68\% containment of the relative error on the energy. Energy bias : relative bias on the energy. \textit{bottom right}: Ratio of sensitivity vs energy.}
\label{fig:lhfitallperf}
\end{figure*}

During an observation the gamma-ray signal is contaminated by electrons and hadrons. In order to optimise the discovery potential of the telescope, we apply data skimming based on a minimum image intensity and a maximum image leakage. The image intensity is the sum of the charge in the camera after cleaning obtained during the Hillas parameter extraction. At very low intensities, the signal is lost in the baseline variability, thus limiting the classification power. The leakage parameter is a measure of how much of the signal falls outside of the camera. It is evaluated as the fraction of the intensity of the cleaned image in the last two rows of the camera pixels. A large value of the leakage can indicate an unreliable reconstruction, for which a small fraction of the real shower image was used.
We optimize the final event selection to achieve the best sensitivity\footnote{\label{note1}Defined as the minimal flux needed in an energy bin to reach a 5$\sigma$ detection with 50h of observations while selecting at least 10 signal events with a signal/background of at least 5\% and with a ratio of acceptance(source region / background only region) of 0.2.} in each energy bin. Five bins per energy decade in log scale are used. We indicate as `gammaness' the output of the RF classifier, showing how gamma-like an image appears and define $\theta$ as the angle between the expected source position and the reconstructed event. Selecting a gammaness maximum corresponds to selecting a true positive rate and the associated false positive rate following from the RF classifier. Consequently, the event selection based on gammaness controls the signal over noise ratio of the selected events. Increasing its value improves the signal over noise ratio and the signal purity at the cost of statistics.
An upper limit on $\theta$ is used in the sensitivity computation to create an ON region containing both signal and background events. Since the gamma event distribution is peaked at the source position, accepting events reconstructed at larger $\theta$ also increases the number of selected signal events at the cost of decreasing the signal over noise ratio. The performance of our method, evaluated on MC is shown on Fig.\ref{fig:lhfitallperf} where the "standard" performance corresponds to the analysis pipeline where the fitting procedure is not performed and the extracted Hillas parameters are used instead. The likelihood method outperforms the standard reconstruction at the lower energies and then shows similar performance above around 100 GeV. 

\section{Application to the data of the Crab Nebula}

Once the performance of the method and the event selection are optimised for sensitivity to a simulated source, the method is applied to observations of the Crab Nebula made by the LST-1 in November 2020. A total of 118 minutes of observations taken with a telescope pointing between $10^\circ$ and $27^\circ$ from the zenith are analysed.

\begin{figure}[ht]
    \centering
    \includegraphics[width = \textwidth]{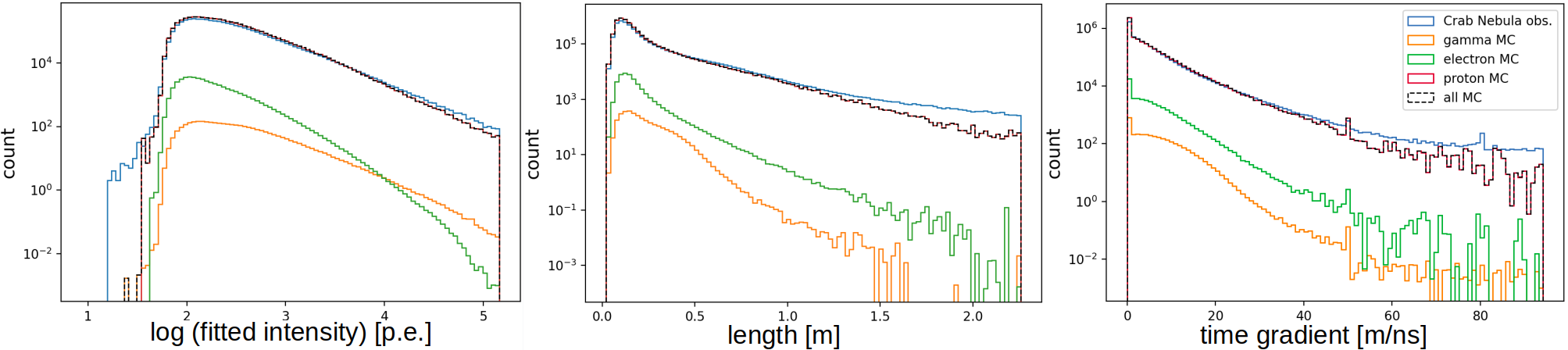}
    \caption{Distribution of a subset of reconstructed image parameters comparing LST-1 data from observations of the Crab Nebula and MC simulation scaled to the observation duration and expected spectra.}
    \label{fig:dl1compar}
\end{figure}

The distribution of parameters extracted during the fitting procedure for a subset of 20 minutes of our data are compared to the predicted distributions from gamma, proton and electron simulations. A good agreement is found after removing events with an Hillas intensity of less than 50 photo-electron (p.e.) (see Fig.\ref{fig:dl1compar}). Such cleaning is required to remove localised excesses from stars in the field of view. This agreement indicates that our MC simulations are a good representation of the data acquired with the LST-1, or at least for the properties to which our method is sensitive.

\begin{figure*}[ht]
\begin{minipage}{0.58\textwidth}
  \centering
  \includegraphics[width=\textwidth]{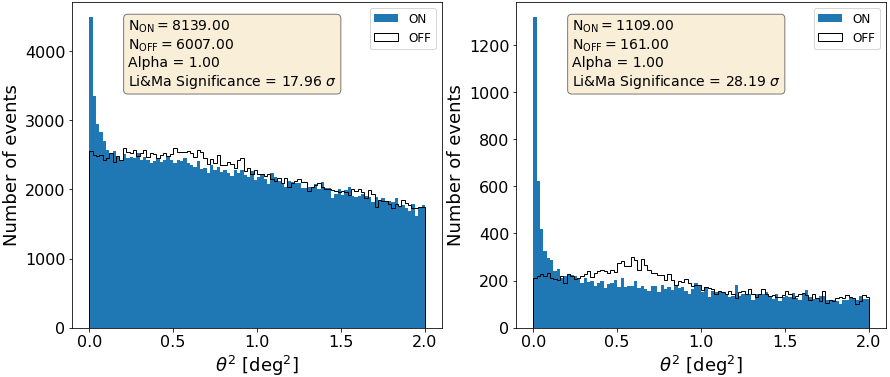}
  \caption{Signal distribution as a function of the angular distance theta from the center of the ON region and average distribution in the $\text{alpha}^{-1}$ OFF regions. \textit{left} : $50 \text{ GeV} < E_{reco} < 25 \text{ TeV}$. \textit{right} : $200 \text{ GeV} < E_{reco} < 2 \text{ TeV}$. }
\label{fig:theta2data}
\end{minipage}
\hfill
\begin{minipage}{0.41\textwidth}
  \centering
  \includegraphics[width=0.89\textwidth]{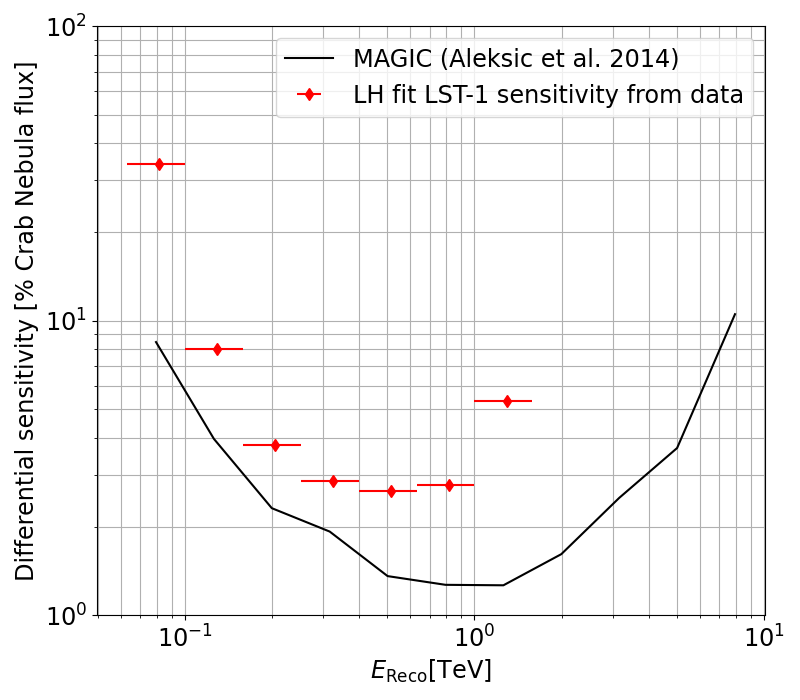}
  \caption{Differential sensitivity\textsuperscript{\ref{note1}} evaluated on Crab Nebulae LST-1 data.}
\label{fig:lhfitsensi}
\end{minipage}
\end{figure*}

The event distribution after selection for the full observation is shown on~Fig.\ref{fig:theta2data}. The background is estimated in an OFF region which sees an excess centered at $\theta^2 = 0.64^\circ$ corresponding to the source position. 
We extract the source significance using equation 17 of Ref.~\cite{1983ApJ...272..317L}. A clear detection above 17$\sigma$ is obtained. Stacking the event distribution at all energies does not fully exploit the detection power of the telescope. Indeed, the high signal/background rate at high energy is diluted in the high accepted background at low energy. This is clearly illustrated by selecting an energy range with better gamma/hadron separation in the right of ~Fig.\ref{fig:theta2data}. 
Extrapolating the observed event rates to 50h, the sensitivity curve in Fig.\ref{fig:lhfitsensi} can be created. It represents the measured sensitivity of the LST-1  at this stage of its commissioning when observing a region of the sky with high NSB and using our reconstruction method. The sensitivity obtained here is similar to the one obtained using a source dependent analysis and the standard reconstruction in Ref.~\cite{physperfICRC2021}.

\begin{figure*}[ht]
  \centering
  \includegraphics[height=4.48cm]{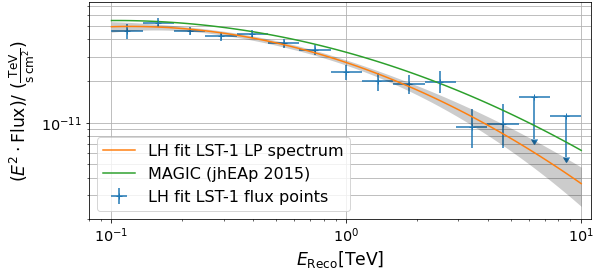}
  \caption{Log parabola fit of the Crab spectrum from our analysis. The reference is from MAGIC~\cite{2015JHEAp...5...30A}.}
\label{fig:spectrumdata}
\end{figure*}

A spectral analysis is performed using a fixed event selection ($\text{gammaness} > 0.7$, $\theta < 0.2$) and gammapy~\cite{gammapy:2017,gammapy:2019}.
The extracted log-parabola spectrum is shown in Fig.\ref{fig:spectrumdata}. It shows a similar spectral shape as the MAGIC log-parabola spectrum but with a lower flux at all energies also seen with the standard reconstruction~\cite{physperfICRC2021}. The fitted model is at $84\pm4\%$ of the MAGIC spectrum flux at 1 TeV.

\section{Conclusion}

A reconstruction method for IACT images exploiting the time-resolved images of the LST camera is introduced. The novelty compared to other likelihood methods lies in the introduction of the shower time development in the image model. The method currently outperforms the standard reconstruction on MC simulation in the lower energies, with a sensitivity up to $\sim$30\% better at 25 GeV and improved angular and energy resolution. The possibility of improvements at higher energy needs to be investigated. Applied to the Crab Nebula observation by the LST-1, the prototype LST of CTA still in commissioning, promising results are obtained. The method is currently based on a simple model, which requires refinements based on a dedicated MC study of images considering spatial asymmetries, dispersion of photon arrival time, detailed multiple photo-electron pulse shape, improved treatment of the detector calibration, an adjustments to the RF feature set. All of these points are being or will be studied to further improve the method.
\vspace{0.5pc}
\newline
\noindent{\small \textbf{Acknowledgements :} We gratefully acknowledge financial support from the agencies and organizations listed here: http://www.cta-observatory.org/consortium\_acknowledgments}

\bibliographystyle{mnras}
{\small
\bibliography{biblio.bib} }

%% Full authors list (ONLY FOR COLLABORATIONS)
\clearpage
\section*{Full Authors List: \Coll}

%\noindent \textbf{Note comment afterwards:} Collaborations have the possibility to provide an authors list in xml format which will be used while generating the DOI entries making the full authors list searchable in databases like Inspire HEP. For instructions please go to icrc2021.desy.de/proceedings or contact us under icrc2021proc@desy.de.\\

\vspace{-0.3pc}
\scriptsize
\noindent
H. Abe$^{1}$,
A. Aguasca$^{2}$,
I. Agudo$^{3}$,
L. A. Antonelli$^{4}$,
C. Aramo$^{5}$,
T.  Armstrong$^{6}$,
M.  Artero$^{7}$,
K. Asano$^{1}$,
H. Ashkar$^{8}$,
P. Aubert$^{9}$
% ...
\\

\vspace{-0.4pc}
\noindent
$^{1}$Institute for Cosmic Ray Research, University of Tokyo.
$^{2}$Departament de Física Quàntica i Astrofísica, Institut de Ciències del Cosmos, Universitat de Barcelona, IEEC-UB.
$^{3}$Instituto de Astrofísica de Andalucía-CSIC.
$^{4}$INAF - Osservatorio Astronomico di Roma.
$^{5}$INFN Sezione di Napoli.
$^{6}$Aix Marseille Univ, CNRS/IN2P3, CPPM.
$^{7}$Institut de Fisica d'Altes Energies (IFAE), The Barcelona Institute of Science and Technology.
$^{8}$IRFU, CEA, Université Paris-Saclay.
$^{9}$LAPP, Univ. Grenoble Alpes, Univ. Savoie Mont Blanc, CNRS-IN2P3, Annecy.
% ...

%first.author$^1$, 
%second.author$^2$, 
%third.author$^3$ % .... more names
%and 
%last.author$^{n}$ \\

%\noindent
%$^1$first.affiliation.
%$^2$second.affiliation. % .... more affiliation
%$^{m}$last.affiliation.

\end{document}